\documentclass[12pt]{article}
\usepackage{pst-all}
\usepackage{pstricks}
\usepackage{pstcol,pst-fill,pst-grad}
\usepackage{amsfonts}
\usepackage{graphicx}
\usepackage{amsmath}
\usepackage{subcaption}

\makeatletter \@addtoreset{equation}{section}

\newcommand{\be}{\begin{equation}}
\newcommand{\ee}{\end{equation}}
\newcommand{\bea}{\begin{eqnarray}}
\newcommand{\eea}{\end{eqnarray}}
\begin{document}

\title{\bf \Large
Dark Energy Effects \\ on Charged and Rotating Black Holes }
\author{   A. Belhaj$^1$,  A. El Balali$^{2}$,  W. El Hadri$^{2}$,
H. El Moumni$^{3,4}$, M. B. Sedra$^{2,5}$
\hspace*{-8pt} \\
%EndAName
{\small $^1$ Equipe des Sciences de la Mati\`ere et du Rayonnement,
ESMAR, D\'{e}partement de Physique }\\ {\small Facult\'e des Sciences, Universit\'e Mohammed V de  Rabat,  Rabat, Morocco} \\
{\small $^{2}$    D\'{e}partement de Physique, LabSIMO,  Facult\'{e}
des Sciences, Universit\'{e} Ibn Tofail }\\{ \small K\'{e}nitra,
Morocco} \\  {\small  $^3$ High
Energy and Astrophysics Laboratory, FSSM,   Cadi Ayyad University}\\
{\small P.O.B. 2390, Marrakesh, Morocco.}\\  {\small  $^4$ LMTI,
Faculty of Sciences, Ibn Zohr University, B.P 8106, Agadir, Morocco}
\\  {\small  $^5$  Facult\'e des Sciences et Techniques, Universit\'e Moulay Isma\"{\i}l, Errachidia, Morocco}
} \maketitle

%\date{}
%\vspace{-3.6em}
%
%\begin{center}
%\textit{$^1$
%\\ [0.5em]
%\end{center}
%
%\vspace{1em}

\begin{abstract}
{\noindent} Using canonical typicality method, we reconsider the
study of dark energy effects on four dimensional black holes.
Concretely, we investigate the  associated  influences on the spectrum of
various black hole backgrounds including the charged and  the rotating
ones. For such  black hole  solutions, we first elaborate analytically the corresponding  radiation
 spectrum, the Hawking temperature and  the dark information.
 Then, we discuss  and analyze  the corresponding findings.
 This work, recovering the results of the Schwarzschield black hole,
 confirms that dark energy can  be considered as a cooling system
 surrounding the black holes providing a colder radiation
  and a slower  Hawking radiation process.

\end{abstract}
\newpage
\tableofcontents

\newpage
\section{Introduction}

From many years, black holes  have been considered  as  important and fascinating
objects, appearing in  various  theories of gravity. Such objects
can provide some insight with respect to the comprehension of
certain aspects of quantum gravity theory\cite{a0,a13}. It  has been
suggested that black holes can  behave as thermodynamic objects  emitting
radiation from the event horizon by exploiting  results of  quantum
field theory  on   curved space-time. This is known by Hawking
radiation, which  has been   explored in the understanding of  black hole thermodynamics
 on non trivial spaces \cite{w1,w2}. Certain  thermodynamical properties
have been investigated for many black holes\cite{w3}.     In AdS  black hole thermodynamics, the cosmological constant $\Lambda$  has been viewed  as a dynamical quantity
associated with a thermodynamical variable. According to \cite{w30}, $\Lambda$   has been interpreted as a thermodynamical pressure
by using the fact that the conjugate thermodynamic variable  is proportional  to a volume. In this way,  the corresponding phase transitions have been extensively
studied in order to support the correspondence between the critical
behaviors of the Van der Waals gas and the charged black holes
\cite{{w4},w5,w6,w7,w8,w9,w10,w11,w12,w13,w130}.

Recently,  certain black holes  have  been  investigated  in the presence
of Dark Energy (DE) considered as one of the hardest puzzles in
modern physics and astronomy  due to  its  exact origin problem
\cite{a1,a2}.   It is recalled that the  dark energy was related to the cosmological
constant and vacuum energy long time ago the observation of the
acceleration of the recent universe. However,  it is still motivated and
supported  by the recent experimental  observations \cite{ade1,ade2}.  Compared to
the fundamental energy density of gravity, the current DE, under this
vision, is very quite small $\rho _{DE}\sim \left( 10^{-3}GeV\right)
^{4}$ \cite{w18,w19}.  In this regard, any attempt associated with
enlightening this tiny magnitude, however dominant energy, is
welcome.   In particular, the  black  holes surrounded  by DE have
received considerable emphasis and their thermodynamics have  been
intensively investigated \cite{Chabab:2017xdw,Li:2014ixn}.  More details on  properties of the black holes in quintessential fields can be found in \cite{q1,anas1,anas2,q2,q3,q4,q5,q6,q7,q8,q9,q10}.

 More
recently, it has been shown that DE affects the radiation spectrum
of black holes.  In the presence of such an energy,  the radiation
spectrum of the Schwarzschild black hole has  been  obtained using
either the statistical mechanical approach or the tunneling approach
\cite{w14,w15,w16,w17,a4}.

The aim of this paper is to contribute to this rich area by
investigating the effect of DE on other  black hole backgrounds
including the charged and the  rotating ones. More precisely, we  first
give the corrected radiation spectrum, the Hawking temperature and the
dark information of such black  holes. Then, we  discuss and analyze
the corresponding results.  Among others, the  Schwarzschild
black hole results  are  recovered by sending the extra
parameters to zero. It has been argued that DE can be considered  as
a cooling system surrounding the black holes providing
 a colder radiation  and a slower  Hawking radiation process.
  In  the present  paper,  we will use dimensionless units in which one has  $\hbar=c=k_{B}=G=1$.

The outline of this paper is as follows. In section 2, we  present
a concise  review  on the canonical approach,  which has been developed,   in order to derive the
radiation spectrum of the Schwarzschild black hole. Section 3 deals
with the spectrum of the  Reissner-Nordstrom (RN) black hole. Section 4  concerns results associated with
  rotating black holes. In section 5,
 we discuss and analyze the corresponding results. The last section is devoted
  to conclusions and open questions.

\section{Black hole radiation spectrum }
In this section, we give a review on the black hole spectrum physics \cite{A}. In particular,
 we discuss the radiation spectrum by recalling the  explored computation
method.

\subsection{Radiation spectrum}
It is shown that, inside the black holes, the quantum
fluctuations create and annihilate a large   number of particle
pairs near the horizon. In this way,   the positive energy particles
escape through tunneling from  the black hole,  inside  region where
the Hawking radiation occurs. Once  such particles cross the black
hole horizon, the statistical  energy distribution can be described
using  the radiation spectrum. It has been remarked that  the
determination  of the  Hawking radiation deserves
deeper    investigations.  The
radiation spectrum,   found by Hawking,  obeys the thermal
distribution \cite{key-sw1,key-sw2}. Alternatively, a statistical
mechanical approach has been also discovered \cite{a4,a5}, bringing
the same results as the quantum tunneling method based on  particle
dynamics \cite{QT1,QT2,QT3,QT4,QT5,QT6,QT7}.

In what follows, we recall such an approach  being  based on
canonical typicality \cite{w15,w16,w17}, or the micro-canonical
hypothesis. It is recalled  that a   black hole $B$  could involve a
mass $M$, a charge $Q$, and an angular momentum $J$. The  corresponding
density matrix  reads as
\begin{equation}
\rho_{{B}}=\sum_{i}\frac{1}{\Omega_{{B}}\left(M,Q,J\right)}\left|M,Q,J\right\rangle
_{i}\left\langle M,Q,J\right|, \label{eq1}
\end{equation}
where $\left|M,Q,J\right\rangle _{i}$ denotes  the $i$th eigenstate of
$B$. However,  $\Omega_{B}\left(M,Q,J\right)$  is  the
associated microstate number. This  expression can   be obtained by
assuming that  a universe $\textrm{U}$  is composed by two subsets
$B$ and $O$ being identified with  the black hole and the
environment, respectively. The black hole evaporation  provides  two
other subsystems which are the radiation field $R$ and the remaining
black hole $B^{\prime} \equiv B-R$. In this way,  the  density
matrix reduces to
\begin{equation}
\rho_{{B}}=\sum_{r,b'} \frac{1}{\Omega_{{B}}\left(E\right)}
\left|r,b'\right\rangle \left\langle r,b'\right|.
\end{equation}
It is noted that $\left|r\right\rangle $ and $\left|b'\right\rangle
$ are the eigenstates of $R$  and $B^{\prime}$ with  eigen-energies
$E_{r}$ and $E_{b'}$, respectively. Here,   $\Omega_{B}\left(E\right)$ is
 the number of the microstates with energy $E=E_{r}+E_{b'}$  required   by the   energy
 conservation principal. The calculation associated with
$B^{\prime}$ shows that  the reduced density matrix of ${R}$  takes the following form
\begin{equation}
\rho_{{R}}=\textrm{Tr}{}_{{B}'}\left(\rho_{{B}}\right)=\sum_{r}\frac{\Omega_{{B}'}
\left(E-E_{r}\right)}{\Omega_{{B}}\left(E\right)}\left|r\right\rangle \left\langle r\right|,
\label{eqRdensity}
\end{equation}
with $\Omega_{B^{\prime}}\left(E-E_{r}\right)$ is the number of
micro-states of $B^{\prime}$ with energy $E-E_{r}$.  Moreover,  the
micro-canonical distribution allows  one  to express  the entropies as
 $S_{{B}}=\ln\Omega_{{B}}\left(E\right)$ and
$S_{{B}^{\prime}}=\ln\Omega_{{B}^{\prime}}\left(E -
E_{r}\right)$. Using these expressions,  the above density matrix
 can be rewritten as
\begin{equation}
\rho_{{R}}=\sum_{r}\textrm{e}^{-\Delta
S_{{BB}'}\left(E_{r},E\right)}\left|r\right\rangle
\left\langle r\right|=\sum_{\omega, \, q, \, j} p \left(
\omega, q, j, M, Q, J \right) \left|\omega, q, j\right\rangle
\left\langle \omega, q, j\right|, \label{eqLdensity}
\end{equation}
where  $p \left( \omega, q, j, M, Q, J \right)$
represents  the distribution probability of the radiation given by
\begin{equation}
p \left( \omega, q, j, M, Q, J \right)= \textrm{e}^{-\Delta S_{{BB}'}\left( \omega, q, j, M, Q, J \right)}.
\label{eqp}
\end{equation}
It is noted that  $ \left|\omega, q, j\right\rangle  $ is the eigenstate of the
radiation $\textrm{R}$, with  mass $\omega$, charge $q$ and angular
momentum $j$. In the black hole physics,  $\Delta
S_{\textrm{BB}'}\left( \omega, q, j, M, Q, J \right)$ denotes  the
entropy difference between ${B}$
  and ${B}^{\prime}$  taking  the following form
\begin{equation}
\Delta S_{{BB}'}\left( E_{r},E\right)=\Delta S_{{BB}'}\left( \omega, q, j, M, Q, J \right)= S_{{B}}
\left( M, Q, J \right)- S_{{B}'}\left( M-\omega, Q-q, J-j \right).
\label{eqS}
\end{equation}
Using \eqref{eqLdensity}-\eqref{eqS}, we can compute the radiation
spectrum  in terms of  $p\left( \omega, q, j, M, Q, J \right)$. To
do so, it is worth noting that  the entropy can be obtained  by using  the
Bekenstrein-Hawking formula
\begin{equation}
S_{BH}\left(M,Q,J\right)=\frac{A_{H}\left(M,Q,J\right)}{4}=\pi R_{H}^{2}\left(M,Q,J\right),
\label{eq:BH entropy-1}
\end{equation}
where $A_{H}\left(M,Q,J\right)$ and $R_{H}\left(M,Q,J\right)$ are
the area and the radius of the black hole, respectively. In this way, the
radiation spectrum,  given in  \eqref{eqp},  can be rewritten as
\begin{equation}
p\left(\omega,q,j,M,Q,J\right)=\textrm{e}^{-\pi\left[R_{H}^{2}\left(M,Q,J\right)-R_{H}^{2}\left(M-\omega,Q-q,J-j\right)\right]}.
\label{eqpf}
\end{equation}
\subsection{Schwarzschield black hole spectrum}
Here, we examine the  Schwarzshield
black hole which will  be considered as a reference model, in order to  check the validity of the present computations.
\subsubsection{Radiation spectrum}
We start by deriving the radiation spectrum for the Schwarzschild
black hole in the existence of the DE using   a
statistical mechanical approach.  To do so,  we  first   express the
horizon radius in terms of  the  mass function using the associated
metric. The latter is given by
\begin{equation}
\textrm{d}s^{2}=f(r) \textrm{d}t^{2}-f^{-1}(r) \textrm{d}r^{2}-
r^{2}\left(\textrm{d}\theta^2+\sin^{2}\theta\textrm{d}\phi^{2}\right).
\end{equation}
The   DE  effect appears in   such a  metric via the $f(r)$
function given by
\begin{equation}
f(r)=1-\frac{2M}{r}-2\left(\frac{r_{o}}{r}\right)^{3\omega_{q}+1},
\label{Schwazshieldm}
\end{equation}
where $r_o$ is a scale factor depending on
the cosmological constant $\Lambda$,  and  where $\omega_{q} $
indicates the state parameter restricted to lie in $-1 < \omega_{q}
<-1/3$.
    It is noted that   the $f(r)$ function  can be obtained by  considering  first   the Einstein
equations for a  physical contribution corresponding to a   static spherically symmetric quintessence field surrounding a
black hole \cite{q1,anas1,anas2}. Then, one should  solve such equations exactly for a specific choice of  free parameters,
  associated  with  energy-omentum tensor of the quintessence noted  by $T_{\mu \nu}$.   For a static state involving  a  spherical symmetry,  it is noted  that   one can  write  a general form
of space-time  components of  such a tensor  as follows
\begin{equation}
T_{t}^{\,t}=A(r), \quad T_{t}^{\,j}=0, \quad T_{i}^{\,j}=-C(r)  \, r_{i} r^{j} + B(r)  \, \delta_{i}^{j}.
\end{equation}
It is emphasized that, for  an  isotropic state,  one can get
\begin{equation}
\left\langle T_{i}^{\,j} \right\rangle =D(r) \, \delta_{i}^{j}, \quad D(r) =-\frac{1}{3} C(r)  \, r^{2}+B(r).
\end{equation}
In the presence of  the quintessence field, one has
\begin{equation}
D(r)=-\omega_{q} \, A(r).
\end{equation}
Fixing the state parameter $\omega_{q}$,  one can obtain  the explicit  expression of the function $D(r)$  representing  the pressure in terms of  the density $A(r)$. Then, it is possible to  parameterize the space-time interval of spherically symmetric static gravitational field as follows
\begin{equation}
\textrm{d}s^{2}= e^{\nu} \textrm{d}t^{2}- e^{\tau}  \textrm{d}r^{2}-
r^{2}\left(\textrm{d}\theta^2+\sin^{2}\theta\textrm{d}\phi^{2}\right)
\label{equation}
\end{equation}
where one has  used  $\nu=\nu(r)$ and $\tau=\tau(r)$. Normalizing   the gravitational constant, we can  determine the general expression of the    quintessence tensor  $T_{\mu \nu}$. In this way, its  spatial part is  proportional to the time component with the arbitrary parameter $B$ depending  on the internal structure of the quintessence.   Up to calculations, one gets
\begin{equation}
T_{t}^{\,t}=\rho_{q}(r), \quad T_{i}^{\,j}=\rho_{q}(r)  \, \alpha \left( -(1+3B) \frac{r_{i} r^{j}}{r_{n} r^{n}} + B(r)  \, \delta_{i}^{j} \right).
\end{equation}
Using $\left\langle r_{i} r^{j} \right\rangle=\frac{1}{3} r_{n} r^{n} \delta_{i}^{j} $,  one can obtain
\begin{equation}
\left\langle T_{i}^{\,j} \right\rangle =- \rho_{q}(r)  \,  \frac{\alpha}{3}  \, \delta_{i}^{j} =-p_{q}(r)\delta_{i}^{j}.
\end{equation}
Thus, the pressure takes the  following form
\begin{equation}
p_{q}=\omega_{q} \, \rho_{q}, \quad \omega_{q}=\frac{\alpha}{3} .
\end{equation}
Using  the metric components $T_{t}^{\,t}=T_{r}^{\,r}$,   and the following equations
\begin{align}
& 2T_{t}^{\,t}=-e^{-\tau} \left(\frac{1}{r^{2}}-\frac{\tau '}{r} \right) + \frac{1}{r^{2}}, \\
& 2T_{r}^{\,r}=-e^{-\tau} \left(\frac{1}{r^{2}}+\frac{\nu '}{r} \right) + \frac{1}{r^{2}},
\end{align}
we can define a principle of linearity and additivity given by $\nu + \tau=0$, with $v^{'}=dv/dr$.  Taking  $\tau=-ln(1+f)$,  one  gets a  linear differential form of Einstein equations  as follows
\begin{align}
& T_{t}^{\,t}= T_{r}^{\,r}=-\frac{1}{r^{2}} \left( f +r f' \right),\label{TENSOR1} \\
& T_{\theta}^{\,\theta}=T_{\phi}^{\,\phi}=-\frac{1}{4r} \left( 2f' +r f'' \right)\label{TENSOR2}.
\end{align}
Using the condition of additivity and linearity,  we can  fixe the free parameter of the energy momentum tensor $B$. Indeed, one considers
\begin{equation}
B=-\frac{3\omega_{q}+1}{6\omega_{q}},
\end{equation}
leading  to
\begin{align}
& T_{t}^{\,t}= T_{r}^{\,r}=\rho_{q},\label{TENSOR3}\\
& T_{\theta}^{\,\theta}=T_{\phi}^{\,\phi}=-\frac{1}{2}\rho_{q} \left( 3\omega+1 \right).
\label{TENSOR4}
\end{align}
Using equations (\ref{TENSOR1}-\ref{TENSOR2}-\ref{TENSOR3}-\ref{TENSOR4}),  we can obtain
\begin{equation}
\left(3\omega+1\right)f+3\left(\omega+1\right)r \, f'+r^{2} \,f''=0.
\end{equation}
The solutions can be given as follows
\begin{equation}
f_{q}=-\frac{c}{r^{(3\omega_{q}+1)}}, \quad f_{BH}=-\frac{r_{g}}{r},
\end{equation}
with $c$ and $r_{g}=2M$ are normalization factors. Since we deal with here   three types of black holes,  it  is useful  to generalize the metric  form  using a free parameter $\omega_{n}$  being  different from the quintessence state parameter $\omega_{q}$. It has been shown that  the form of the exact spherically symmetric solutions of Einstein equations can  have the following form
\begin{equation}
\begin{aligned}
 \textrm{d}s^{2}= & \left[ 1-\frac{2M}{r}-\sum_{n} \left( \frac{r_{n}}{r} \right)^{3\omega_{n}+1} \right] \textrm{d}t^{2}  \\
& -\left[ 1-\frac{2M}{r}-\sum_{n} \left( \frac{r_{n}}{r} \right)^{3\omega_{n}+1} \right]^{-1} \textrm{d}r^{2} -r^{2}\left(\textrm{d}\theta^2+\sin^{2}\theta\textrm{d}\phi^{2}\right)
\end{aligned}
\end{equation}
where $r_{n}$ are now  dimensional normalization constants. Taking $\omega_{n}=1/3$ for instance,  with the corresponding normalization constant being equal to $-Q^{2}$, one     can get the Reissner-Nordstorm black hole function $f(r)$ given by
\begin{equation}
f(r)=1-\frac{2M}{r}+\frac{Q^{2}}{r^{2}}.
\end{equation}
To obtain the de Sitter space one can take $\omega_{n}=-1$, where the normalization constant is $\ell^{2}$. For a  particular contribution of quintessence,  one can write
\begin{equation}
f(r)=1-\frac{2M}{r}+\frac{Q^{2}}{r^{2}}-\left( \frac{r_{q}}{r} \right) ^{3\omega_{q}+1}=1-\frac{2M}{r}+\frac{Q^{2}}{r^{2}}-2 \left( \frac{r_{0}}{r} \right) ^{3\omega_{q}+1},
\end{equation}
  where $r_{q}^{3\omega_{q}+1}=2r_{0}^{3\omega_{q}+1}=c$. Taking $Q=0$, however, we obtain the Schwarzshield black hole with quintessence. In this way,  the  metric function is given by  \eqref{Schwazshieldm}. It is known that  the de-Sitter Schwarzschield black hole has a metric function given by
\begin{equation}
f(r)=1-\frac{2M}{r}-\frac{r^{2}}{\ell^{2}},
\end{equation}
which is noting but  the function $f(r)$ of  the Schwarzschield black hole surrounded by quintessence with  $\omega_{q}=-1$. The associated function is  written as
\begin{equation}
f(r)=1-\frac{2M}{r}-2\frac{r^{2}}{r_{0}^{2}}.
\end{equation}
It turns out that  can relate the de-Sitter parameter $\ell$ to $r_{0}$ by $\frac{r_{0}}{\sqrt{2}}=\ell=\sqrt{\frac{3}{\Lambda}}$   providing  the expression $r_{0}=\sqrt{\frac{6}{\Lambda}}$.
To determine the horizon radius of the Schwarzshield black holes surrounded by the  quintessence, we   can exploit  the metric function given in \eqref{Schwazshieldm}, and   then resolve the following equation
\begin{equation}
f(R_{H})=1-\frac{2M}{R_{H}}-2 \left( \frac{6}{\Lambda R_{H}^{2}} \right) ^{\frac{3\omega_{q}+1}{2}}=0.
\end{equation}
Assuming that the  quintessence  affects slightly the geometry of the black hole, we can resolve such an  equation perturbatively by taking $R_{H}=R_{H}^{0}+\delta$, where $\delta$ represents the modification or perturbation generated by the presence of quintessence.  Indeed, the above equation takes the following form
\begin{equation}
 1-\frac{2M}{R_{H}^{0}+\delta}-2 \left( \frac{6}{ \Lambda \left(  R_{H}^{0}+\delta \right)^{2}} \right) ^{\frac{3\omega_{q}+1}{2}}=0.
 \end{equation}
A simple  factorization gives  \begin{equation}
 1-\frac{2M}{R_{H}^{0}\left( 1+ \frac{\delta}{R_{H}^{0}} \right)}-2 \left( \frac{6}{  \Lambda \left( R_{H}^{0}\left( 1+ \frac{\delta}{R_{H}^{0}} \right) \right)^{2}} \right) ^{\frac{3\omega_{q}+1}{2}}=0. \label{dem0}
\end{equation}
Using  the fact that $\frac{\delta}{R_{H}^{0}} \ll 1 $, one can write
 \begin{equation} 1-\frac{2M}{2M}\left( 1+ \frac{\delta}{2M} \right)^{-1}-2 \left( \frac{6}{  \Lambda \left( 2M  \right)^{2}} \right) ^{\frac{3\omega_{q}+1}{2}}. \label{dem1}  \end{equation}
Moreover, applying   $ \left( 1+ \frac{\delta}{R_{H}^{0}} \right)^{-1} = \left( 1- \frac{\delta}{R_{H}^{0}} \right) $, one gets
  \begin{equation}  1-\left( 1- \frac{\delta}{2M} \right) -2 \left( \frac{6}{  \Lambda \left( 2M  \right)^{2}} \right) ^{\frac{3\omega_{q}+1}{2}} \label{dem2}  \end{equation}
 Taking $\xi=-\frac{3\omega_{q}+1}{2}$, one obtains
\begin{align}
 \frac{\delta}{2M}  -2 \left( \frac{6}{  \Lambda \left( 2M  \right)^{2}} \right) ^{-\xi} =0, \end{align} leading to \begin{equation}
\delta_{Sch}=2 \left( 2M \right)^{2\xi+1} \left(\frac{\Lambda}{6}\right)^{\xi}.
\end{equation}
Note in passing that   this  method will be exploited, trough this paper,   to determine the horizon loci of   black holes
treated in the present work.  Roughly, the corresponding  horizon radius reads as
\begin{equation}
R_{H}^{Sch}=2M+4M^{2\xi+1}F_{Sch},\label{eq:A}
\end{equation}
where $\xi=-(3\omega_{q}+1)/2 $ belonging to the interval $\left[0,1\right]$ and where $F_{Sch}$ is a function given by
\begin{equation}
F_{Sch}=F_{Sch}\left(\xi,\Lambda\right)=\left(\frac{2\Lambda}{3}\right)^{\xi}. \label{Fsch}
\end{equation}
Replacing \eqref{eq:A} in \eqref{eqpf}, we obtain the corrected radiation spectrum in the presence of DE
\begin{equation}
p\left(\omega,M\right)=\textrm{exp}\left\{ -8\pi
M\left[1+4\left(\xi+1\right)M^{2\xi}F_{Sch}\right]\omega+4\pi\left[1+4\left(\xi+1\right)\left(2\xi+1\right)M^{2\xi}F_{Sch}\right]\omega^{2}\right\}.
\label{C}
\end{equation}
 Turning off the effect of  DE $F_{Sch} \rightarrow 0$ ($ \Lambda \rightarrow 0$), this  probability
reduces
to
\begin{equation}
p\left(\omega,M\right)=\exp\left[-8\pi\omega\left(M-\omega/2\right)\right]
\label{PW}
\end{equation}
which is the  Parikh-Wilczek (P-W) spectrum  reported in \cite{X}.
\subsubsection{Temperature}
Analyzing  the Hawking radiation as tunneling, Parikh and Wilczek have  noticed that the radiation spectrum can not  be strictly thermal [52]. However, they  have given a correction of the probability of emission. In the absence of  quintessence,  the  above  probability of emission is   called the non thermal radiation. In fact, it is not easy to  derive the Hawking temperature from this radiation. However, one  should  deal with  only the thermal part which is achieved  by  considering  large black holes ($M \gg 1$).
Since $\omega$ is the mass of the radiated particle, it is obviously much smaller than $M$.  In this way,  the above relation reduces to
\begin{equation}
p(\omega,M)=e^{\left(-8 \pi M \omega \right)}=e^{- \beta \omega}.
\end{equation}
where $\beta$ is the inverse of Hawking temperature. Thus, the temperature of Schwarzshield Black hole is
\begin{equation}
T=\frac{1}{8 \pi M}.
\end{equation}
 To derive the temperature from the thermal radiation spectrum, similar analysis is needed in the rest of the paper.

For  the radiation spectrum in the presence of DE, we
discuss the evaporation process.  as stated above for  a black
hole with a large mass ($M\gg1$), the non-thermal contribution
of the radiation spectrum can be evinced. Then,  the  approximated
thermal distribution probability becomes now as
\begin{equation}
p\left(\omega,M\right)=\exp\left\{ -8\pi
M\left[1+4\left(\xi+1\right)M^{2\xi}F_{Sch}\right]\omega\right\}
.\label{D}
\end{equation}
In the presence of DE, further calculations associated with \eqref{D} gives
$\beta_{H}^{Sch}$ leading to  the following expression of the  Hawking temperature
\begin{equation}
T_{H}^{Sch}=\frac{1}{8\pi
M}\left[1-4\left(\xi+1\right)M^{2\xi}\left(\frac{2\Lambda}{3}\right)^{\xi}\right].\label{eq:Tsch}
\end{equation}
Introducing a modification factor
\begin{equation}
\lambda_{Sch}=4\left(\xi+1\right)M^{2\xi}\left(\frac{2\Lambda}{3}\right)^{\xi},
\end{equation}
this  Hawking temperature takes the following form
\begin{equation}
T_{H}^{Sch}=T_{H}^{Sch_{0}}(1-\lambda_{Sch}),
\end{equation}
where $T_{H}^{Sch_{0}}$ indicates now   the Hawking
temperature in the absence of  DE  given by
\begin{equation}T_{H}^{Sch_{0}}=1/\left(8\pi M\right)\end{equation}

Having discussed a simple black hole, we are going  to consider
generic  models by introducing other  physical parameters including
charge and  the spin.  Concretely, we investigate the effect of DE on
the corresponding  black hole solutions.
\section{Spectrum of Reissner-Nordstrom black hole}
\subsection{Radiation spectrum}
In this section,  we would like to  derive the radiation spectrum of
the Reissner-Nordstrom (RN) black hole in the existence of DE using
the above discussed  statistical mechanical approach. Like the
previous discussions, we first  determine  the horizon radius. Turning on
the effect of the DE, the $f(r)$ metric function, for the RN black hole, is
written as
\begin{equation}
f(r)=1-\frac{2M}{r}- \frac{Q^{2}}{r^{2}} -2\left(\frac{r_{o}}{r}\right)^{3\omega_{q}+1},
\end{equation}
where $M$ is the mass of the black hole and $Q$ is its
charge\cite{darkenergymetric}. In this way, the  horizon
radius  can be  obtained by solving the constraint $ f
\left(R_{H}\right)=0 $. Taking $r_{0}= \sqrt{\frac{6}{\Lambda}}$, the
calculation gives
\begin{equation}
1-\frac{2M}{R_{H}} - \frac{Q^{2}}{R_{H}^{2}}-2\left(\frac{6}{\Lambda
R_{H}^{2}}\right)^{\frac{3\omega_{q}+1}{2}}=0.\label{eq:rh}
\end{equation}
The solution of \eqref{eq:rh} can be put as
$R_{H}^{RN}=R_{H}^{0}+\delta_{RN}$. The first term
$R_{H}^{0}=M+\sqrt{M^{2}-Q^{2}}$ is the horizon radius of the RN
black hole  without DE. While, the second one carries
information on the effect of DE. For such a black hole, this term  is given by
\begin{equation}
\delta_{RN}=\frac{\left( M + \sqrt{M^{2} - Q^{2}}
\right)^{2\xi+2}}{\sqrt{M^{2}-Q^{2}}} \left(\frac{\Lambda}{6}
\right)^{\xi}.
\end{equation}
Thus, the associated horizon radius  reads as
\begin{equation}
R_{H}^{RN}=M+\sqrt{M^{2}-Q^{2}}+\frac{\left( M + \sqrt{M^{2} -
Q^{2}} \right)^{2\xi+2}}{\sqrt{M^{2}-Q^{2}}} F_{RN}, \label{eq:R}
\end{equation}
where $F_{RN}$ is a function  depending on $\xi$ and $\Lambda$.   For such a black hole, it  is given by
\begin{equation}
F_{RN}=F_{RN}\left(\xi,\Lambda\right)=\left(\frac{\Lambda}{6}\right)^{\xi},
\label{fRN}
\end{equation}
obtained by  keeping only  the first order of $F_{RN}$. Taking
$Q=0$, we recover the horizon radius of  the Schwarzschield black
hole given in \eqref{eq:A}. It is noted that the modified RN black
hole horizon radius is quite complicated. The corresponding
calculations need  relevant approximations. It follows from
\eqref{eq:R} that DE increases the horizon radius. This can be
understood from the fact that DE produces the opposite effect being
repulsion, contrary  to the gravitation produces an attraction force
for regular objects. This could  explain  the expansion of the
Universe. Obviously,  the extension of  the black hole horizons should be
considered  as a consequence of the presence of DE. Indeed, the
repulsion caused by
DE which appears naturally from \eqref{eq:R} leads to a  larger black hole.\\
The influences of  DE on the black hole radiation can be
discussed  through the statistical mechanical method  introduced in
the previous section. Toward the determination of the radiation
spectrum, we use several simplifications and approximations.
Substituting \eqref{eq:R} into \eqref{eqpf}, for instance,  we find
the term
\begin{equation}
\begin{aligned}
{R^{RN}_{H}}^2\left(M,Q\right)- &{R^{RN}_{H}}^2\left(M-\omega,Q-q\right) =\left(M+\sqrt{M^{2}-Q^{2}}\right)^{2}-\left(M-\omega+\sqrt{\left(M-\omega\right)^{2}-\left(Q-q\right)^{2}}\right)^{2} \\
&+2F_{RN} \left\lbrace \frac{\left(M+\sqrt{M^{2}-Q^{2}}\right)^{2\xi+3}}{\sqrt{M^{2}-Q^{2}}}-\frac{\left(M-\omega+\sqrt{\left(M-\omega\right)^{2}
-\left(Q-q\right)^{2}}\right)^{2\xi+3}}{\sqrt{\left(M-\omega\right)^{2}-\left(Q-q\right)^{2}}} \right\rbrace
\end{aligned}
\end{equation}
up to the first order in $F_{RN}$. An examination shows that  the  first term can be simplified.
However, the second one is stuck. The problem here is the square
root functions and the denominators. As we are dealing with objects
that have big masses (many times the solar mass $10^{30}$ ), we can
consider the case $Q \ll M$. The second  term,  which will be denoted by
$\alpha_{DE} $, can be rewritten as
\begin{equation}
\begin{aligned}
\alpha_{DE} &= 4F_{RN} \left\lbrace \left(M+\sqrt{M^{2}-Q^{2}}\right)^{2\xi+2} \, \frac{\left(M+\sqrt{M^{2}-Q^{2}}\right)}{2\sqrt{M^{2}-Q^{2}}} \right. \\
&\qquad \qquad \left. - \left(M-\omega+\sqrt{\left(M-\omega\right)^{2}-\left(Q-q\right)^{2}}\right)^{2\xi+2}\,  \frac{\left(M-\omega+\sqrt{\left(M-\omega\right)^{2}-\left(Q-q\right)^{2}}\right)}{2\sqrt{\left(M-\omega\right)^{2}-\left(Q-q\right)^{2}}} \right\rbrace.
\end{aligned}
\end{equation}
Using the approximations $M \simeq \sqrt{M^{2}-Q^{2}}$ and $M-\omega \simeq \sqrt{\left(M-\omega\right)^{2}-\left(Q-q\right)^{2}}$, $\alpha_{DE}$
can be expressed as
\begin{equation}
\alpha_{DE} \simeq 4F_{RN} \left\lbrace \left(M+\sqrt{M^{2}-Q^{2}}\right)^{2\xi+2}-\left(M-\omega+\sqrt{\left(M-\omega\right)^{2}-\left(Q-q\right)^{2}}\right)^{2\xi+2} \right\rbrace.
\end{equation}
To handle these square root functions,  we shall use  the following
limit  $\sqrt{1-x}\simeq 1-\frac{x}{2}$. After calculations, the corrected
radiation spectrum of the RN black hole reads as
\begin{equation}
p\left(\omega,q,M,Q\right)= \textrm{exp}\left\{-\pi \left( \alpha_{RN} + \alpha_{DE} \right) \right\},
\label{eq:p}
\end{equation}
with
\begin{equation*}
\begin{aligned}
 \alpha_{RN} & = 2M \sqrt{M^{2} -Q^{2} } + 2\omega \left( 2M-\omega \right) - q \left( 2Q-q \right) -2\left(M-\omega \right) \sqrt{\left(M-\omega \right)^2 -\left(Q-q \right)^2 }, \\
 \alpha_{DE} & = 4 \left( \xi+1 \right)
  \left\lbrace 2\left( 2 \omega + q \, \frac{q-2Q}{2\left(M- \omega\right)} \right)  \left(2M-\frac{Q^{2}}{2
  \left( M- \omega \right)}\right)^{2\xi+1} \right. \\
& \qquad \qquad \left. - \left(2 \xi+1 \right) \left( 2 \omega + q \, \frac{q-2Q}{2\left(M- \omega\right)} \right)^{2}
 \left(2M-\frac{Q^{2}}{2\left( M- \omega \right)}\right)^{2\xi} \right\rbrace F_{RN}
\end{aligned}
\end{equation*}
where we have kept the second order of $\omega$ and  $q$.  The
higher orders are omitted.  It is noted that the first term
$\textrm{exp}\left\{-\pi \alpha_{RN} \right\}$ concerns the
radiation spectrum without DE.  However, the  extra  factor
$\alpha_{DE}$ is associated with  the DE correction. This reveals
that the radiation spectrum depends on $\xi$ and $\Lambda$. When the
effect of DE vanishes, the above probability would reduce to
\begin{equation}
p\left(\omega,q,M,Q\right)=\textrm{e}^{-\pi\left[ \left(M+\sqrt{M^{2}-Q^{2}} \right)^{2}- \left(M-\omega+\sqrt{\left( M - \omega
\right)^{2}-\left( Q - q \right)^{2}} \right)^{2}\right]}=\textrm{e}^{-\pi \alpha_{RN}}.
\end{equation}
 It is worth noting that we can check the validity of the main  results by comparing them to the Schwarzschield black hole  case. Taking $Q=0$ and $q=0$, we recover \eqref{C}
which matches perfectly with the results given in \cite{a4}.  It is noted  that the  black hole radiation spectrum  can be investigated
using an  alternative  method  based on the  so-called quantum tunneling,
which will not be  discussed here.   For the absence of DE, we can
use, however, the results  obtained in \cite{X,Y}  to verify  the
validity of  the present results  concerning the  absence of DE.

\subsection{Evaporation process  the  RN black hole }
Here,  we reconsider  the Hawking temperature of  the RN black holes
with the  DE corrections.  It is recalled that such a temperature is
\begin{equation}
T_{H}^{RN_{0}}=\frac{\sqrt{M^{2}-Q^{2} }}{2 \pi \left( M + \sqrt{M^{2}-Q^{2}
} \right)^{2} }.
\label{Twijdane}
\end{equation}
For a black hole with a large mass ($M\gg1$), the non-thermal part of
the radiation spectrum can be ignored. In this way,  the
approximated thermal distribution probability  reads as
\begin{equation}
\begin{aligned}
p\left(\omega,q,M,Q\right)& =\exp\left( -\pi \left\lbrace 2M \left[\sqrt{M^{2}-Q^{2}} -
\sqrt{ \left( M- \omega \right)^{2}-\left(Q-q\right)^{2}} \right] \right. \right. \\
& +2 \omega \left[ 2M + \sqrt{ \left( M- \omega \right)^{2}-\left(Q-q\right)^{2}} + 8 \left( \xi+1\right)  F_{RN}
\left(2M - \frac{Q^{2}}{2\left(M-\omega \right)} \right)^{2\xi+1}  \right] \\
& \left. \left. +2q \left[ -Q + 8 \left( \xi+1\right)  F_{RN} \left( \frac{q-2Q}{2 \left(M- \omega \right)} \right)
 \left(2M - \frac{Q^{2}}{2\left(M-\omega \right)} \right)^{2\xi+1} \right] \right\rbrace \right).
\end{aligned}
\label{eq:thermal}
\end{equation}
It is observed that,  when the effect of the DE vanishes
($F_{RN}\rightarrow0$), this distribution  reduces to the ordinary
Hawking radiation spectrum
\begin{equation}
\begin{aligned}
p\left(\omega,q,M,Q\right)= \exp & \left( -\pi \left[  2M \sqrt{M^{2}-Q^{2}}  +4 M \omega -2 Qq \right.\right. \\
& \left. \left. -2 \left( M-\omega \right)\sqrt{ \left( M- \omega \right)^{2}-\left(Q-q\right)^{2}} \right] \right).
\end{aligned}
\label{eq:pp}
\end{equation}
In what follows, we  discuss  such  a spectrum in terms of DE.
\subsubsection{Without DE}
Here,  we deal with  two  cases.  The first one  concerns the
 neutral radiated particles $(q=0)$  while the second
 one is associated with  charged radiated particles   ($q \neq 0$). In the first
case, \eqref{eq:pp} becomes
\begin{equation}
p\left(\omega,q,M,Q\right)=\exp\left( -\pi \left[ 2M
\sqrt{M^{2}-Q^{2}}+4 M \omega -2 \left( M-\omega \right) \sqrt{
\left( M- \omega \right)^{2}-Q^{2}} \right] \right).
\label{eq3.13}
\end{equation}
Using appropriate  approximations, we get
\begin{equation}
p\left(\omega,q,M,Q\right)=\exp\left( -\beta_{H}^{RN_{0}} \omega \right)=\exp
\left( -2 \pi \frac{\left(M+\sqrt{M^{2}-Q^{2}}\right)^{2}}{\sqrt{M^{2}-Q^{2}}} \omega\right),
\label{eq3.14}
\end{equation}
where $\beta_{H}^{RN_{0}}$, being  the inverse of the temperature in the
absence of DE,  reads  as
\begin{equation}
\beta_{H}^{RN_{0}}=2 \pi \frac{\left(M+\sqrt{M^{2}-Q^{2}}\right)^{2}}
{\sqrt{M^{2}-Q^{2}}}.
\end{equation}
In the second case ($q \neq 0$,   however, similar approximations give
\begin{equation}
\begin{aligned}
p\left(\omega,q,M,Q\right)& =\exp\left( -\beta_{H}^{RN_{0}} \left( \omega -\omega_{0} \right)\right),
\\
& =\exp \left( -2 \pi
\frac{\left(M+\sqrt{M^{2}-Q^{2}}\right)^{2}}{\sqrt{M^{2}-Q^{2}}}\left(
\omega -\omega_{0} \right)\right)
\end{aligned}
\end{equation}
where we have  used $ \omega_{0} = q \, \frac{Q}{M+\sqrt{M^{2}-Q^{2}}} $. It is noted that these results
match perfectly with
 the quantum tunneling method developed in \cite{X,Y}.
\subsubsection{With DE}  In this part, we discuss generic situations associated with \eqref{eq:thermal}. Taking $q=0$ ($ \omega_{0}=0$), we
find
\begin{equation}
\begin{aligned}
p\left(\omega,q,M,Q\right)& =\exp\left( -\pi \left\lbrace 2M \left[\sqrt{M^{2}-Q^{2}} -
\sqrt{ \left( M- \omega \right)^{2}-Q^{2}} \right] \right. \right. \\
& \left. \left. +2 \omega \left[ 2M + \sqrt{ \left( M- \omega \right)^{2}-Q^{2}} + 8 \left( \xi+1\right)
 F_{RN} \left(2M - \frac{Q^{2}}{2\left(M-\omega \right)} \right)^{2\xi+1} \right]  \right\rbrace \right).
\end{aligned}
\end{equation}
Using the expression of $\beta_{H}^{RN_{0}}$ and leaving the added term by DE unchanged, we obtain
\begin{equation}
p\left(\omega,q,M,Q\right) = \exp \left(-\beta_{H}^{RN_{0}} \omega-16 \pi \omega \left( \xi+1\right) F_{RN}
 \left(2M - \frac{Q^{2}}{2\left(M-\omega \right)} \right)^{2\xi+1} \right).
\end{equation}
Factorizing by $\beta_{H}^{RN_{0}} \omega$, this can be put as
\begin{equation}
p\left(\omega,q,M,Q\right) =\exp\left( -\beta_{H}^{RN_{0}} \omega  \left\lbrace 1 + 8 \left( \xi+1\right) F_{RN}
 \left(2M - \frac{Q^{2}}{2\left(M-\omega \right)} \right)^{2\xi+1} \frac{\sqrt{M^{2}-Q^{2}}}{ \left(M+\sqrt{M^{2}-Q^{2}}\right)^{2}} \right\rbrace\right).
\end{equation}
Considering  the situation where the mass of the particle ($\omega$)
is very small compared to the black hole for only one power of
$\left(2M - \frac{Q^{2}}{2\left(M-\omega \right)} \right)$, we can
find
\begin{equation}
p\left(\omega,q,M,Q\right) \simeq \exp\left( -\beta_{H}^{RN_{0}} \omega  \left\lbrace 1 + 4 \left( \xi+1\right)  F_{RN} \left(2M - \frac{Q^{2}}{2\left(M-\omega \right)}
\right)^{2\xi}\frac{2\sqrt{M^{2}-Q^{2}}}{ \left(M+\sqrt{M^{2}-Q^{2}}\right)} \right\rbrace\right).
\end{equation}
Taking $Q \ll M$, we can write $M\simeq \sqrt{M^{2}-Q^{2}} $. In this way, the probability distribution takes the following form
\begin{equation}
\begin{aligned}
p\left(\omega,q,M,Q\right)=& \exp\left( -2 \pi \frac{\left(M+\sqrt{M^{2}-Q^{2}}\right)^{2}}{\sqrt{M^{2}-Q^{2}}}  \right. \times \\
& \left. \left\lbrace 1 + 4 \left( \xi +1 \right) F_{RN} \left(2M - \frac{Q^{2}}{2\left(M-\omega\right)} \right)^{2\xi} \right\rbrace \omega  \right).
\end{aligned}
\end{equation}
Using this equation, we get
\begin{equation}
\beta_{H}^{RN} = \frac{ 2 \pi \left(M+\sqrt{M^{2}-Q^{2}}\right)^{2}}{\sqrt{M^{2}-Q^{2}}} \left\lbrace 1
 + 4\left( \xi +1 \right)  F_{RN} \left(2M - \frac{Q^{2}}{2\left(M-\omega\right)} \right)^{2\xi} \right\rbrace.
\label{eq:beta}
\end{equation}
It is  expected that these calculation steps and approximations can
be exploited   for other black hole solutions including Kerr and
Kerr-Newmann  ones. The  temperature of the RN black hole, in the
presence of DE, reads as
\begin{equation}
T_{H}^{RN} = \frac{\sqrt{M^{2}-Q^{2}}}{2 \pi \left(M+\sqrt{M^{2}-Q^{2}}\right)^{2}}
\left\lbrace 1 - 4 \left( \xi +1 \right)  F_{RN} \left(2M - \frac{Q^{2}}{2\left(M-\omega\right)} \right)^{2\xi} \right\rbrace.
\end{equation}
Using \eqref{Twijdane},  the Hawking temperature,  with the
existence of DE, takes the following form
\begin{equation}
T_{H}^{RN} = T_{H}^{RN_{0}} \left\lbrace 1 - 4 \left( \xi +1 \right) F_{RN} \left(2M - \frac{Q^{2}}{2\left(M-\omega\right)} \right)^{2\xi} \right\rbrace,
\label{eq:T}
\end{equation}
where we only kept the first order of $F_{RN}$. Introducing a
modification factor
\begin{equation}
\lambda_{RN}=4\left(\xi+1\right)F_{RN}\left(2M - \frac{Q^{2}}{2\left(M-\omega\right)} \right)^{2\xi},
\end{equation}
the Hawking temperature can be further put  as
\begin{equation}
T_{H}^{RN}=T_{H}^{RN_{0}}(1-\lambda_{RN})\label{eq:HT}.
\end{equation}
For $q \neq 0$, we obtain, however,
\begin{equation}
\begin{aligned}
p\left(\omega,q,M,Q\right)& =\exp\left( -2 \pi \frac{\left(M+\sqrt{M^{2}-Q^{2}}\right)^{2}}
{\sqrt{M^{2}-Q^{2}}} \cdot \right. \\
& \left. \left\lbrace 1 + 4 \left( \xi +1 \right)  F_{RN} \left(2M -
\frac{Q^{2}} {2\left(M-\omega\right)} \right)^{2\xi} \eta
\right\rbrace \left( \omega - \omega_{0} \right)  \right)
\end{aligned}
\end{equation}
where $\eta$ is a function depending on the RN black hole parameters given by
\begin{equation}
 \eta=\frac{1-Qq/\left(2M\omega\right)}{1-Qq/\left(\omega
\left(M-\frac{Q^{2}}{2M}\right)\right)} - \frac{M}{2M-\frac{Qq}{\omega}-\frac{Q^{2}}{2M}}.
\end{equation}
Further calculations give
\begin{equation}
\beta_{H}^{RN} = \beta_{H}^{RN_{0}} \left\lbrace 1 + 4 \left( \xi +1 \right) F_{RN} \left(2M - \frac{Q^{2}}{2\left(M-\omega\right)}
 \right)^{2\xi} \eta \right\rbrace.
\label{eq:beta2}
\end{equation}
Then, the temperature is
\begin{equation}
T_{H}^{RN}=T_{H}^{RN_{0}} \left\lbrace 1 - 4 \left( \xi +1 \right) F_{RN} \left(2M - \frac{Q^{2}}{2\left(M-\omega\right)} \right)^{2\xi} \eta \right\rbrace.
\label{eq:beta222}
\end{equation}
It is remarked from \eqref{eq:beta} and \eqref{eq:beta222} that DE makes the Hawking radiation colder.
Taking $Q=0$ and $q=0$, we recover the Schwarzschild black hole case given by
\begin{equation}
\beta_{H}^{Sch}=8\pi M\left[1+4\left(\xi+1\right)M^{2\xi}F_{Sch}\right].
\label{Bsch}
\end{equation}
This produces  the  ordinary  temperature appearing in \eqref{eq:Tsch}.
\subsection{Dark information added by DE }
To investigate  the information carried by  the Hawking radiation, we
should use  the  Von-Neumann entropy $ S_{R}= - Tr \left(\rho_{R}
\ln\rho_{R}\right) $ of $ R $ which reads as
\begin{equation}
S_{R}= S_{B}- S\left(B^{\prime}|R\right).
\end{equation}
Here,     $  S\left(B^{\prime}|R\right)$ is given by
\begin{equation}
S\left(B^{\prime}|R\right)=  \sum\limits_{r} e^{- \Delta
S_{BB^{\prime}} \left(E_{r}, E\right)} \,
S_{B^{\prime}}\left(E-E_{r}\right),
\end{equation}
which is  the conditional entropy of ${B}^{\prime}$.
 This shows that there is a correlation between $R$ and ${B}^{\prime}$.  Taking  low energy limit,
  ie., $E_{r} << E$, ${B}^{\prime}$ is approximated  by keeping the first order of $ E_{r}$. It is noted that
$ E\left(B^{\prime}|R\right)\simeq S_{B^{\prime}}\left(E_{B^{\prime}}\right)$. Here, $ E_{B^{\prime}}= E-E_{R}$ and
   $E_{R}=\sum\limits_{r} \exp \left(-\Delta S_{BB^{\prime}} \left(E_{r},E\right)\right) \, E_{r}$ are the internal energies
   of $ {B}^{\prime}$ and $R$, respectively. In other words, the correlation between $R$ and $B$
    could be ignored when the energy of $R$ is much smaller than that  the one of  $B$. This  implies  the existence of the
     thermal radiation discovered by Hawking \cite{key-sw1,key-sw2}. It turns out that  the thermal spectrum of the black hole
 radiation is due to the ignorance of the correlation information between the black hole and its radiation. Concretely,
this is also the primary cause of the black hole information loss \cite{a5}.

It is recalled that the  non-thermal radiation has been shown  to be
the origin of the information correlation between the emissions
being radiated out from the  horizon of the black hole.  Consider the influence of  DE on the dark information of neutral
particles $(q=0)$. Indeed, the mutual information for two emissions
$\omega_{a}$ and $\omega_{b}$ can be
 written as
\begin{equation}
I\left(a,b\right)=\sum\limits_{\omega_{a}\omega_{b}} p_{a,b} \ln\left(\dfrac{p_{a,b}}{\left( p_{a} \,p_{b}\right)}\right).
\end{equation}
In the case of the RN black hole, it follows from \eqref{eq:p} that  the distribution probabilities
for two particles $a$ and $b$  read as
\begin{equation}
\begin{aligned}
&p_{a}= p\left(\omega_{a},M\right)= \exp\left(- \beta_{H}^{RN} \, \omega_{a}+ \Gamma_{H}^{RN} \,  \omega_{a}^{2}\right), \\
&p_{b}= p\left(\omega_{b},M\right)= \exp\left(- \beta_{H}^{RN} \,
\omega_{b}+ \Gamma_{H}^{RN} \, \omega_{b}^{2}\right).
\end{aligned}
\label{eq:w}
\end{equation}
It is  noted that $\beta_{H}^{RN}$ is the inverse radiation temperature and $\Gamma_{H}^{RN}$
 given by
\begin{equation}
\Gamma_{H}^{RN}= 4 \pi \left( 1+4 \left( \xi+1\right) \left(2 \xi+1\right)  F_{RN} \left(2M-\dfrac{Q^{2}}{2\left(M-\omega\right)}\right)^{2\xi} \right).
\end{equation}
Then, the  joint probability  takes the following  form
\begin{equation}
p_{a,b}=  p\left(\omega_{a},M\right) \cdot
p\left(\omega_{b},M^{\prime}\right)=\exp\left(- \beta_{H}^{RN}
\left(\omega_{a}+\omega_{b}\right)+ \Gamma_{H}^{RN}
\left(\omega_{a}+\omega_{b}\right)^{2} \right), \label{eq:z}
\end{equation}
where  $M^{\prime}= M - \omega$. Using \eqref{eq:w} and
\eqref{eq:z}, the dark information becomes
\begin{equation}
I^{RN} \left( a,b\right)= 2 \Gamma_{H}^{RN} E_{a} E_{b},
\end{equation}
where  the internal energies  of the particles $a$ and $b$ are given
by
\begin{equation}
E_{a}= \sum\limits_{\omega_{a}=0}^{\omega_{a}=M}
p\left(\omega_{a},M\right) \cdot \omega_{a}, \qquad E_{b}=
\sum\limits_{\omega_{b}=0}^{\omega_{b}=M^{\prime}}
p\left(\omega_{b},M^{\prime} \right) \cdot \omega_{b}.
\end{equation}
Thus, the  dark information for the RN black hole  can be  expressed
as
\begin{equation}
I^{RN} \left(a,b\right)=8\pi\left[1+4 \left(2\xi+1\right)\left(\xi+1\right) F_{RN} \left(2M-\dfrac{Q^{2}}{2\left(M-\omega\right)}\right)^{2\xi}\right]E_{a}E_{b}.
\end{equation}
At this level,  we wish to add some comments on this  expression.
The first comment concerns the fact   that the mutual information
between the emissions is proportional to their internal energies.
Taking into account information correlation between the radiation
particles, the present  result   can  confirm  that the RN black
hole information is not lost.  The second comment is that one can
recover the results of \cite{a4}. Indeed, taking  $Q=0$,  we find
\begin{equation}
I_{Sch}\left(a,b\right)=8\pi\left[1+4\left(\xi+1\right)\left(2\xi+1\right)M^{2\xi}F_{Sch}\right]E_{a}E_{b}.
\label{Isch}
\end{equation}

\section{DE effects on rotating black holes}
In this section, we investigate  DE influences on the rotating black
hole solutions. Precisely, we establish the associated spectrum.
\subsection{Kerr black hole}
First, we consider  the  Kerr black hole which  is an uncharged
black hole  which rotates    about a central axis. However,  it has only
   mass and angular momentum  considered as a   possible generalization of the Schwarzschild black
  hole. This  black hole, having  spinning structure,  involves  two surfaces:  the horizon, region
   from which no signal can escape and the Ergosphere which rotates so fast. Moreover,  the
   relevant feature of such a black  hole solution  is that it describes all black holes without electric charges. It has been proposed that the
    Kerr  solution can be considered as a  general description of
      astrophysical balck holes \cite{A1}.
\subsubsection{Radiation spectrum}
According to \cite{a8,a6},  the metric of the  Kerr black hole,  in
the presence of quintessence field,  is given by
\begin{equation}
    \begin{aligned}
        \mathrm{d}s^{2}& =-\left(1-\frac{2M \, r + c r^{1-3\omega_{q}}}{\Sigma}\right)
        \mathrm{d}t^{2}+\frac{\Sigma}{\Delta}\mathrm{d}r^{2} -
        2a \sin^{2}\theta\left(\frac{2M \, r + c r^{1-3\omega_{q}}}{\Sigma}\right)
        \mathrm{d}\phi \, \mathrm{d}t + \Sigma \mathrm{d}\theta^{2} \\
                       &+\sin^{2}\theta\left\lbrace r^{2} + a^{2}+a^{2}\sin^{2}\theta
                        \left(\frac{2M \, r + c r^{1-3\omega_{q}}}{\Sigma}\right) \right\rbrace\mathrm{d}\phi^{2},
    \end{aligned}
\end{equation}
where  $ \Sigma=r^{2}+a^{2} \cos^{2}\theta $ and  $a=\frac{J}{M}$.
$J$  and $ M $ are the spin and  the mass of such   a  black hole,
respectively. It is recalled that $c$ is the quintessential
parameter  indicating the intensity of the quintessence energy
contribution  linked to the black hole metric  via the relation
\begin{equation}c=2 \left( \frac{6}{\Lambda} \right)
^{\frac{3\omega_{q}+1}{2}}.
\end{equation}
In this black hole solution, we have
\begin{equation}
\Delta=r^{2}-2Mr+a^{2}-cr^{1-3\omega_{q}}.
\end{equation}
To  get the outer horizon, we  should solve the  equation  $\Delta=0$.
Without DE,  we get  $R^{0}_{H}=M+\sqrt{M^{2}-a^{2}} $. In
the presence of DE,  the  solution of  $\Delta=0$
 can be perturbatively  put like $R^K_{H}=R^{0}_{H}+\delta$.   Taking
$\delta/R^{0}_{H} \ll 1$, we get
\begin{equation}
\delta_{K}=\frac{\left( M + \sqrt{M^{2} - a^{2}} \right)^{2\xi+2}}{\sqrt{M^{2}-a^{2}}} \left(\frac{\Lambda}{6} \right)^{\xi}.
\end{equation}
Thus,  the  corresponding horizon radius takes the following form
\begin{equation}
R_{H}^{K}=M+\sqrt{M^{2}-a^{2}}+\frac{\left( M + \sqrt{M^{2} - a^{2}}
\right)^{2\xi+2}}{\sqrt{M^{2}-a^{2}}} F_{K}. \label{eq:RK}
\end{equation}
In this black hole solution, the function $F_{K}$ is given by
\begin{equation}
F_{K}=F_{K}\left(\xi,\Lambda\right)=\left(\frac{\Lambda}{6}\right)^{\xi},
\label{FK}
\end{equation}
where we have only kept the first order of $F_{K}$.  Taking  $0<
\mid a \mid < M$, $\Delta=0$ involves two solutions  representing
the inner and outer horizons.\\ Using the corrected horizon radius,
we can  write down  the distribution probability of a particle with
mass $\omega$ and  spin $j$. Substituting  the horizon radius
\eqref{eq:RK} into \eqref{eqpf}, we can obtain the corrected
radiation spectrum of the  Kerr black hole. Here, the rotation rate
$a$ must  be fixed  in order to guarantee the conservation of the
angular momentum and the energy. In this way, we should  fix the
total mass and the
 total angular momentum and allow  the black hole mass and the angular momentum  to fluctuate \cite{a8}.
  When  a  particle  of spin $j=\omega \cdot a$  escapes out of the black
   hole, the angular momentum becomes $J= \left( M- \omega \right) \cdot a$.
Using similar technics exploited in the RN black hole and
substituting \eqref{eq:RK} in \eqref{eqpf}, we find
   \begin{equation}
\begin{aligned}
R^{2}_{K}\left(M,J\right)- &R^{2}_{K}\left(M-\omega,J-j\right) =\left(M+\sqrt{M^{2}-a^{2}}\right)^{2}-\left(M-\omega+\sqrt{\left(M-\omega\right)^{2}-a^{2}}\right)^{2} \\
&+2F_{K} \left\lbrace
\frac{\left(M+\sqrt{M^{2}-a^{2}}\right)^{2\xi+3}}{\sqrt{M^{2}-a^{2}}}
-\frac{\left(M-\omega+\sqrt{\left(M-\omega\right)^{2}-a^{2}}\right)^{2\xi+3}}{\sqrt{\left(M-\omega\right)^{2}-a^{2}}}
\right\rbrace,
\end{aligned}
\end{equation}
up to the first order of $F_{K}$. An examination shows difficulties
in dealing
 with such an expression due to the square root and the difference in the
 denominators appearing
  in the seconde term. To evince this difficulty, we proceed in the same way as we have done for the RN black hole. Concretely,
   we assume that the mass is bigger than the spin. In the slow rotation rate regime, this   is considered  as a  physical assumption. Roughly, we get
\begin{equation}
\begin{aligned}
R^{2}_{K}\left(M,J\right)- &R^{2}_{K}\left(M-\omega,J-j\right) =\left(M+\sqrt{M^{2}-a^{2}}\right)^{2}-\left(M-\omega+\sqrt{\left(M-\omega\right)^{2}-a^{2}}\right)^{2} \\
&+4F_{K} \left\lbrace \left(M+\sqrt{M^{2}-a^{2}}\right)^{2\xi+2}-\left(M-\omega+\sqrt{\left(M-\omega\right)^{2}-a^{2}}\right)^{2\xi+2} \right\rbrace.
\end{aligned}
\end{equation}
Using $\sqrt{1-x}\simeq 1-\frac{x}{2}$, the final distribution probability reads as
\begin{equation}
    \begin{aligned}
        p\left(\omega,j,M,J\right)= & \exp \left[-\pi \left\lbrace 2M \sqrt{M^{2}-a^{2}}-2\left(M-\omega\right)\sqrt{\left(M-\omega\right)^{2}-a^{2}}
         \right. \right. \\
        & + 4 \omega \left[ M+ 4\left( \xi + 1 \right) F_{K} \left(2M-\frac{a^{2}}{2\left(M-\omega\right)}\right)^{2\xi+1} \right] \\
        &\left. \left. -2\omega^{2}\left[ 1 + 8 \left( \xi + 1 \right)\left(2 \xi + 1 \right) F_{K} \left(2M-\frac{a^{2}}{2\left(M-\omega\right)}
        \right)^{2\xi}  \right]  \right\rbrace \right],
    \end{aligned}
\label{eq:pkerr}
\end{equation}
where we have kept the second order of $\omega$, and the higher
orders are omitted.  It is  observed clearly  the dependence of the
radiation spectrum on $\xi$ and $\Lambda$.   It is noted that the
Schwarzschield radiation spectrum can be  recovered  by sending
$a$ to zero.
\subsubsection{Evaporation process of Kerr black hole}
We move now to  discuss   the modified Hawking temperature  for  the
Kerr black hole in the presence of DE. For the  ordinary Kerr black
hole, one has
\begin{equation}
T_{H}^{K_{0}}= \frac{\sqrt{M^{2}-a^{2} }}{4 \pi
\left( M^{2} +M \sqrt{M^{2}-a^{2} } \right)}.
\label{ker100}
\end{equation} In the case of black hole with a large mass, i.e.,
$M\gg1$, the non-thermal part of the radiation spectrum can be
ignored. Thus, the approximated thermal distribution probability
takes the following form
\begin{equation}
    \begin{aligned}
        p\left(\omega,j,M,J\right)= & \exp \left[-\pi \left\lbrace 2M
         \sqrt{M^{2}-a^{2}}-2\left(M-\omega\right)\sqrt{\left(M-\omega\right)^{2}-a^{2}} \right. \right. \\
        &\left. \left. + 4 \omega \left[ M+ 4\left( \xi + 1 \right)  F_{K} \left(2M-\frac{a^{2}}{2\left(M-\omega\right)}
        \right)^{2\xi+1} \right]  \right\rbrace \right].
    \end{aligned}
\label{eq:thermalK}
\end{equation}
It is remarked that  when the effect of the DE vanishes,
assured by $F_{K} \rightarrow 0$, such a   distribution  reduces to
the Hawking radiation spectrum
\begin{equation}
p\left(\omega,j,M,J\right)=  \exp \left[-\pi \left\lbrace 4M\omega + 2M \sqrt{M^{2}-a^{2}}-2\left(M-\omega\right)\sqrt{\left(M-\omega\right)^{2}-a^{2}}
\right\rbrace \right].
\label{eq:ppk}
\end{equation}
Using \eqref{eq:ppk} and  an appropriate  approximation,  we
obtain
\begin{equation}
p\left(\omega,j,M,J\right)=\exp\left( -\beta_{H}^{K_{0}} \omega \right)=\exp \left( -4 \pi
 \frac{\left(M^{2}+M\sqrt{M^{2}-a^{2}}\right)}{\sqrt{M^{2}-a^{2}}} \omega\right),
\end{equation}
where $\beta_{H}^{K_{0}}$ reads as
\begin{equation}
\beta_{H}^{K_{0}}=4 \pi \frac{\left(M^{2}+M\sqrt{M^{2}-a^{2}}\right)}{\sqrt{M^{2}-a^{2}}} .
\end{equation}
It is noted  that this  agrees with the result of \cite{a7}
obtained using quantum tunneling approach.  In the presence of DE,
however,  we find
\begin{equation}
\begin{aligned}
p\left(\omega,j,M,J\right)= & \exp \left\lbrace \frac{-4 \pi\left(M^{2}+M\sqrt{M^{2}-a^{2}}\right)}{\sqrt{M^{2}-a^{2}}} \right. \\
&\qquad \qquad \left. \left( 1 + 4\left(\xi+1\right) F_{K} \left(2M-\frac{a^{2}}{2\left(M-\omega\right)}\right)^{2\xi}\left(1-\frac{a^{2}}{2M^{2}}\right)\right)
\omega\right\rbrace.
\end{aligned}
\end{equation}
This  gives
\begin{equation}
\beta_{H}^{K}=\frac{4 \pi\left(M^{2}+M\sqrt{M^{2}-a^{2}}\right)}{\sqrt{M^{2}-a^{2}}} \left( 1 + 4\left(\xi+1\right)F_{K}
\left(2M-\frac{a^{2}}{2\left(M-\omega\right)}\right)^{2\xi}\left(1-\frac{a^{2}}{2M^{2}}\right)\right).
\label{eq:betaK}
\end{equation}
Thus,  the corresponding  Hawking temperature  reads as
\begin{equation}
T_{H}^{K}=\frac{\sqrt{M^{2}-a^{2}}}{4 \pi\left(M^{2}+M\sqrt{M^{2}-a^{2}}\right)}
 \left( 1 - 4\left(\xi+1\right) F_{K} \left(2M-\frac{a^{2}}{2\left(M-\omega\right)}\right)^{2\xi}\left(1-\frac{a^{2}}{2M^{2}}\right)\right).
\end{equation}
Using \eqref{ker100},  the Hawking temperature  can be rewritten as
\begin{equation}
T_{H}^{K}=T_{H}^{K_{0}} \left( 1 - 4\left(\xi+1\right)F_{K}
\left(2M-\frac{a^{2}}{2\left(M-\omega\right)}\right)^{2\xi}\left(1-\frac{a^{2}}{2M^{2}}\right)\right).
\end{equation}
Introducing a modification factor
\begin{equation}
\lambda_{K}=4 \left(\xi+1\right) F_{K} \left(2M-\frac{a^{2}}{2\left(M-\omega\right)}\right)^{2\xi}\left(1-\frac{a^{2}}{2M^{2}}\right),
\end{equation}
the Hawking temperature can be further simplified as
\begin{equation}
T_{H}^{K}=T_{H}^{0}(1-\lambda_{K}).\label{eq:HTK}
\end{equation}
It is worth noting  that \eqref{eq:betaK} confirms the
previous results proposing that  DE makes the Hawking radiation
colder.

\subsubsection{Dark information added by DE }
To   extract  dark information  associated with the Kerr  black hole,  we
can   use the  analysis  adopted  for the previous  black solutions.
This is  achieved naively by   a direct calculation.  Indeed, the corresponding   dark information  takes the following  form
  \begin{equation}
I_K\left(a,b\right)=8\pi\left[1+4\left(2\xi+1\right)\left(\xi+1\right)
F_{K}
\left(2M-\dfrac{a^{2}}{2\left(M-\omega\right)}\right)^{2\xi}\right]E_{a}E_{b}.
\end{equation}
This finding matches perfectly with the fact   that the black hole
information is not lost, when  we consider  information
 correlations between radiated particles.

\subsection{Kerr Newmann black hole}
In this subsection,  we  discuss the  Kerr-Newmann black hole  involving
three hairs $ (M, Q, J)$. In this way, one might exploit  the
technical calculations  used in the investigation of   RN and Kerr
black holes. In  what follows, we give only the main results.
\subsubsection{Radiation spectrum}
According to  \cite{a10,a11},   the line element, of  such a black hole solution, can be written as
\begin{equation}
\begin{aligned}
ds^{2}&= -\left(1-\frac{2Mr-Q^{2}}{\Sigma}\right) dt^{2}+\frac{\Sigma}{\Delta}dr^{2}-2a \sin^{2} \theta
\left(\frac{2Mr-Q^{2}}{\Sigma}\right) \, d\phi \, dt+ \Sigma d\theta^{2} \\
& \qquad +\sin^{2}\theta \left[r^{2}+a^{2}+ a^{2} \, \sin^{2} \left(\frac{2Mr-Q^{2}}{\Sigma}\right)\right] d\phi^{2}.
\end{aligned}
\end{equation}
In the presence of quintessence field,  this expression  becomes
\begin{equation}
\begin{aligned}
ds^{2}&= -\left(1-\frac{2Mr-Q^{2}+c r^{1-3\omega_{q}}}{\Sigma}\right) dt^{2}+\frac{\Sigma}{\Delta}dr^{2}-2a \sin^{2}
 \theta \left(\frac{2Mr-Q^{2}+c r^{1-3\omega_{q}}}{\Sigma}\right) \, d\phi \, dt\\
& \qquad + \Sigma d\theta^{2}  +\sin^{2}\theta \left[r^{2}+a^{2}+ a^{2} \, \sin^{2} \left(\frac{2Mr-Q^{2}+c r^{1-3\omega_{q}}}{\Sigma}\right)\right] d\phi^{2},
\end{aligned}
\end{equation}
where
\begin{equation}
\Sigma \equiv r^{2}+ a^{2} \cos ^{2}\theta,\qquad  \Delta = r^{2}-2Mr+a^{2}+Q^{2}-2r^{2} \, \left(\frac{6}{\Lambda r^{2}}\right)^{\frac{3 \omega_{q}+1}{2}}.
\end{equation}
As the previous cases,  the horizon radius can be obtained from  the
above  equation. Using the previous analysis and  taking the horizon
radius of the ordinary  Kerr-Newmann black hole
 \begin{equation}
R_{H}^{KN_{0}}= M+ \sqrt{M^{2}-Q^{2}-a^{2}},
\end{equation}
we can obtain
\begin{equation}
\delta_{KN}=\frac{\left( M + \sqrt{M^{2}-Q^{2}- a^{2}} \right)^{2\xi+2}}{\sqrt{M^{2}-Q^{2}-a^{2}}} \left(\frac{\Lambda}{6} \right)^{\xi}.
\end{equation}
Then, the associated horizon radius is given by
\begin{equation}
R_{H}^{KN}=R_{H}^{KN_{0}}+\frac{\left( M + \sqrt{M^{2}-Q^{2} - a^{2}}
\right)^{2\xi+2}}{\sqrt{M^{2}-Q^{2}-a^{2}}}
F_{KN}\left(\xi,\Lambda\right). \label{eq:RKN}
\end{equation}
Here, the function $F_{KN}$  reads as
\begin{equation}
F_{KN}=F_{KN}\left(\xi,\Lambda\right)=\left(\frac{\Lambda}{6}\right)^{\xi},
\label{FKN}
\end{equation}
where we have only kept the first order of $F_{KN}$.  This corrected
horizon radius  allows one  to derive the distribution probability
of a particle with mass $\omega$, charge $q$ and a spin $j$.  Using
the statistical mechanical method and substituting the horizon
radius \eqref{eq:RKN} into \eqref{eqpf},  we obtain then the
corrected radiation spectrum of the  Kerr Newmann black hole. When
the rotation rate $a$ is fixed,  the angular momentum and the energy
are conserved. With the right approximations, the distribution
probability reads as
\begin{equation}
    \begin{aligned}
        p\left(\omega,q,j,M,Q,J\right)& =  \exp \left[-\pi \left\lbrace 2M \sqrt{M^{2}-Q^{2}-a^{2}}-2\left(M-\omega\right)
        \sqrt{\left(M-\omega\right)^{2}-\left(Q-q\right)^{2}-a^{2}}  \right. \right. \\
        &+4M\omega-2\omega^{2}-2Qq+q^{2}  \\
        & + 8 \left( \xi + 1 \right)F_{KN}\left(2M-\frac{Q^{2}+a^{2}}{2\left(M-\omega\right)}\right)^{2\xi+1}\left(2\omega+q\cdot \frac{q-2Q}{2\left(M-\omega\right)}\right) \\
        &\left. \left. -4 \left( 2\xi + 1 \right) \left( \xi + 1 \right) F_{KN} \left(2M-\frac{Q^{2}+a^{2}}
        {2\left(M-\omega\right)}\right)^{2\xi} \left(2\omega+q\cdot \frac{q-2Q}{2\left(M-\omega\right)} \right)^{2} \right\rbrace \right],
    \end{aligned}
\label{eq:pkerrN}
\end{equation}
where we have kept the second order of $\omega$ and $q$, and the
higher orders are ignored. We can clearly  observe  that the
radiation spectrum depends  on $\xi$ and $\Lambda$. Putting $a=0$ and
$Q=0$, we can recover  the Schwarzschield
 radiation spectrum. RN and Kerr
radiation spectrums are also obtained by taking
$a=0$ an  $Q=0$, respectively.

\subsubsection{Evaporation process of Kerr Newmann black hole}
Here, we study the modified Hawking temperature for the Kerr Newmann
black hole in the  presence of DE. It is recalled that  the ordinary temperature
is
\begin{equation}
T_{H}^{KN_{0}}=\frac{\sqrt{M^{2}-a^{2}-Q^{2} }}{4 \pi \left( M^{2} +M
\sqrt{M^{2}-Q^{2}
 -a^{2} }-Q^{2}/2 \right)}.
\label{Anas}
 \end{equation}
For large mass, i.e., $M\gg1$, the non-thermal part of the radiation
spectrum can be ignored. In this way, the approximated thermal
distribution probability is
\begin{equation}
    \begin{aligned}
        p\left(\omega,q,j,M,Q,J\right) = \exp & \left[ -\pi \left\lbrace 2M \sqrt{M^{2}-Q^{2}-a^{2}}-2\left(M-\omega\right)
        \sqrt{\left(M-\omega\right)^{2}-\left(Q-q\right)^{2}-a^{2}}  \right. \right. \\
        &+4M\omega -2Qq  \\
        & \left. \left. +8\left( \xi + 1 \right) F_{KN} \left(2M-\frac{Q^{2}+a^{2}}{2\left(M-\omega\right)} \right)^{2\xi+1}
         \left(2\omega+q\cdot \frac{q-2Q}{2\left(M-\omega\right)}\right) \right\rbrace \right].
    \end{aligned}
\label{eq:thermalKN}
\end{equation}
Considering the limit  $F_{KN} \rightarrow 0$, the above
distribution reduces to the Hawking radiation spectrum
\begin{equation}
\begin{aligned}
p\left(\omega,q,j,M,Q,J\right)=& \exp \left[-\pi \left\lbrace 2M \sqrt{M^{2}-Q^{2}-a^{2}}-2\left(M-\omega\right)
\sqrt{\left(M-\omega\right)^{2}-\left(Q-q\right)^{2}-a^{2}} \right. \right. \\
& \left. \left. \qquad \qquad +4M\omega  -2Qq \right\rbrace \right].
\end{aligned}
\label{eq:ppKN}
\end{equation}
Using \eqref{eq:ppKN} and the right approximation, we get
\begin{equation}
p\left(\omega,q,j,M,Q,J\right)=\exp \left( -\beta_{H}^{KN_{0}} \omega \right)=\exp
 \left( \frac{ -4 \pi\left( M^{2} + M \sqrt{M^{2}-Q^{2} -a^{2} }-\frac{Q^{2}}{2} \right)}{\sqrt{M^{2}-Q^{2}-a^{2}}} \omega\right).
\end{equation}
This result matches  perfectly with  the work  reported in  \cite{a9,a12}
using quantum tunneling approach.  When  DE is present and the
radiated particle is uncharged, we  obtain
\begin{equation}
\begin{aligned}
p\left(\omega,j,M,J\right)= & \exp \left\lbrace \frac{ -4 \pi\left( M^{2} + M \sqrt{M^{2}-Q^{2} -a^{2} }
-\frac{Q^{2}}{2} \right)}{\sqrt{M^{2}-Q^{2}-a^{2}}} \right. \\
&\qquad \qquad \left. \left( 1 + 4\left(\xi+1\right) F_{KN} \left(2M-\frac{Q^{2}+a^{2}}{2\left(M-\omega\right)}
\right)^{2\xi}\left(1-\frac{a^{2}}{2M^{2}}\right)\right) \omega\right\rbrace.
\end{aligned}
\end{equation}
In this way,  $\beta_{H}^{KN}$ takes the form
\begin{equation}
\begin{aligned}
\beta_{H}^{KN}& = \, \frac{4 \pi\left( M^{2} + M \sqrt{M^{2}-Q^{2} -a^{2} }-\frac{Q^{2}}{2} \right)} {\sqrt{M^{2}-Q^{2}-a^{2}}} \cdot \\
&  \left( 1 + 4\left(\xi+1\right) F_{KN} \left(2M-\frac{Q^{2}+a^{2}}{2\left(M-\omega\right)}\right)^{2\xi} \left(1-\frac{a^{2}}{2M^{2}}\right)\right).
\end{aligned}
\label{eq:betaKN}
\end{equation}
Then,  the Hawking temperature is
\begin{equation}
\begin{aligned}
T_{H}^{KN} &=  \frac{\sqrt{M^{2}-Q^{2}-a^{2}}}{4 \pi\left( M^{2} + M \sqrt{M^{2}-Q^{2} -a^{2} }-\frac{Q^{2}}{2}\right)} \\
& \left( 1 - 4\left(\xi+1\right) F_{KN} \left(2M-\frac{Q^{2}+a^{2}}{2\left(M-\omega\right)}\right)^{2\xi} \left(1-\frac{a^{2}}{2M^{2}}\right)\right).
\end{aligned}
\end{equation}
Using \eqref{Anas}, the Hawking temperature can be put as
\begin{equation}
T_{H}^{KN}=T_{H}^{KN_{0}} \left( 1 - 4\left(\xi+1\right)F_{KN} \left(2M-\frac{Q^{2}+a^{2}}{2\left(M-\omega\right)}
\right)^{2\xi}\left(1-\frac{a^{2}}{2M^{2}}\right)\right).
\end{equation}
Introducing a modification factor
\begin{equation}
\lambda_{KN}=4\left(\xi+1\right)F_{KN} \left(2M-\frac{Q^{2}+a^{2}}
{2\left(M-\omega\right)}\right)^{2\xi}\left(1-\frac{a^{2}}{2M^{2}}\right),
\end{equation}
the Hawking temperature can be further simplified as
\begin{equation}
T_{H}^{KN}=T_{H}^{KN_{0}}(1-\lambda_{KN}).\label{eq:HTKN}
\end{equation}
It follows from  \eqref{eq:betaKN}  that  DE makes the Hawking
radiation colder.
\subsubsection{Dark information added by DE}
In the present black
hole solution,  the dark information can be computed using the same
method. However, we give just the obtained calculation. Indeed, we
have
\begin{equation}
I_{KN}\left(a,b\right)=8\pi\left[1+4\left(2\xi+1\right)\left(\xi+1\right)
F_{KN}
 \left(2M-\dfrac{Q^{2}+a^{2}}{2\left(M-\omega\right)}\right)^{2\xi}\right]E_{a}E_{b}.
\end{equation}
 When the  information correlation between the radiation
  particles is taken into account, this   confirms the idea  that
  the black hole information is not lost.
\section{Results and discussions}
In this section,  we present the result of our analysis.    A close inspection shows that DE increases the black hole size.
This can be understood from the horizon radii of Schwarzschield \eqref{eq:A},
RN \eqref{eq:R}, Kerr \eqref{eq:RK} and Kerr Newmann \eqref{eq:RKN} being a consequence of the  repulsion effect. Beside, the function $F_{Sch}$,
  appearing in \eqref{Fsch}, takes its maximum value for the Schwarzschield black hole case  which is $\left(\frac{2\Lambda}{3}\right)^{\xi}$.
  However, this function becomes $\left(\frac{\Lambda}{6}\right)^{\xi}$ for the RN black hole,  kerr black hole or Kerr-Newmann black hole.
  By comparing these functions,  the  influence of DE on the Schwarzschield black hole is more relevant.
It turns out that  the enhancement in the size of the black hole
corresponds to  the loose of temperature. This can be understood
from the fact that  the resulting surface gravity,  which is inversely
proportional to the black hole horizon square, decreases. Thus, the
temperature, being  proportional to the surface gravity,  decreases
as well.  This result matches with the DE cooling mechanism (DECM)
where DE decreases the Hawking temperature in  the space
expansion context.

It is noted that the factor $\lambda_{i}$ where $i=\text{Sch, RN, K,
KN} $ is proportional  to the mass.  This means  that the influence of DE on the Hawking temperature
increases with the black hole mass. Moreover, in the black-body
radiation theory, the lower temperature is associated with   the smaller  radiation
power. Thus,  the evaporation of such black holes will be longer.

\begin{figure} [h]
\begin{center}
\includegraphics[scale=0.55]{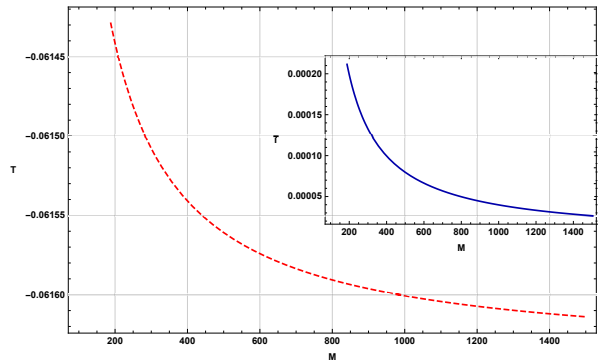}
\caption{Schwarzschield temperature in the presence of DE
(red dashed line) and without DE blue. Note that
$\xi=\frac{1}{2}$.} \label{GSCH}
\end{center}
\end{figure}

To understand such thermodynamical behaviors, we plot the black hole
temperatures in terms of the mass in the presence and the absence of DE. \\
In figure \eqref{GSCH}, we illustrate the case of  the Schwarzschield black hole temperature with DE (red)
and without dark  energy (blue). It follows from  this figure that the DE leads to  less temperature. In this way, the black hole takes more time to evaporate, keeping the behavior associated with the absence of DE.

\begin{figure} [h]
\begin{center}
\includegraphics[scale=0.55]{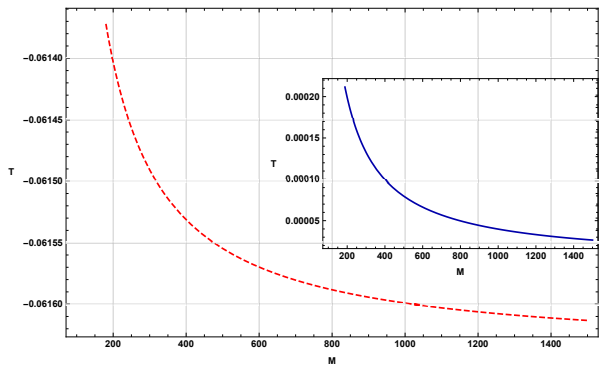}
\caption{Reissner Nordstorm temperature in the presence of dark energy (red dashed line) and without dark energy (blue), where we have used  $Q=10$,
 $\xi=\frac{1}{2}$ and $\omega=-\frac{2}{3}$.}
\label{GRN}
\end{center}
\end{figure}

In figure \eqref{GRN},  we plot the case of  the RN black hole. It is observed the same behavior as the Schwarzschield case.

To evince the extremal case ($Q=M$), we should take only the values of $M$ verifying  $Q<M$. As expected, the temperature decreases in the presence of DE providing a  long time evaporation.

\begin{figure}[h]
\begin{center}
\includegraphics[scale=0.55]{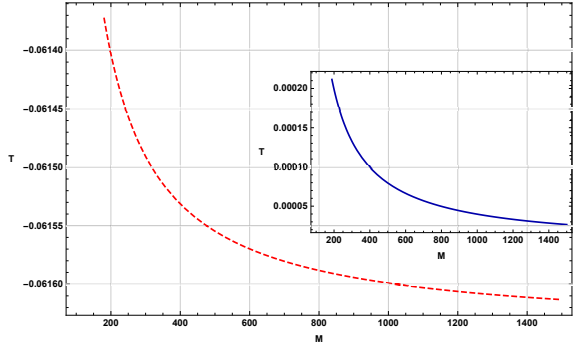}
\caption{Kerr temperature in the presence of dark energy (red dashed line) and without dark energy (blue), with $a=10$, $\xi=\frac{1}{2}$ and $\omega=-\frac{2}{3}$.}
\label{GK}
\end{center}
\end{figure}

The case of the Kerr black hole is presented in figure \eqref{GK}.
With or without DE,  it is noted similar behaviors as the
Schwarzschield case. Avoiding the extremal case  and taking $a<M$,
we find the same results as the previous ones.

\begin{figure}[h]
\begin{center}
\includegraphics[scale=0.55]{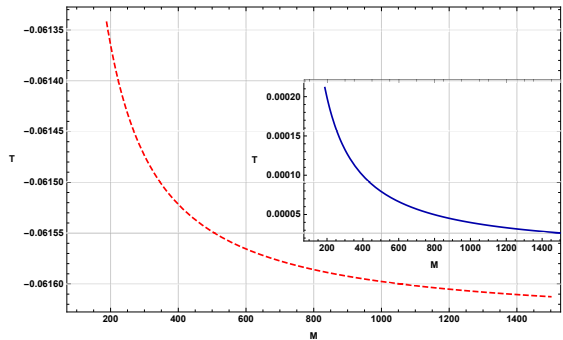}
\caption{Kerr-Newmann temperature in the presence of dark energy (red dashed line) and without dark energy (blue). With $Q=10$,  $a=10$, $\xi=\frac{1}{2}$ and $\omega=-\frac{2}{3}$.}
\label{GKN}
\end{center}
\end{figure}

In figure \eqref{GKN}, we plot the Kerr-Newman  case.  It is observed
the same behavior as the Schwarzschield black hole when DE is
absente. It is recalled that the extremal case is constrained by
$a^{2}+Q^{2}=M^{2}$  representing a circle in the black hole moduli
space. Far from such a region, we recover  the same  results as the
previous black holes.

\section{Conclusion and open questions}

In this paper, we have  investigated  the effect of DE on  charged
and rotating  black holes.  Using a statistical method, we   have
presented  the  associated  radiation spectrum, the Hawking
temperature and dark information of such black holes.  These
results match perfectly with  the ones   corresponding  to the
Schwarzschild black hole. Sending the extra black hole parameters to
zero, we have obtained the spectrum expression of such a black hole.
Then, we have analyzed in some  details  the finding of the present
work. Indeed, it has been confirmed that DE can behave as a cooling
system surrounding the black hole. This produces a colder
radiation and a slower Hawking radiation process. A close examination
on these results reveals that $T_{KN}>T_{K}>T_{RN}>T_{Sch}$. Thus,
this leads to
$R_{Sch}>R_{RN}>R_{K}>R_{KN}$. \\
This work comes up with many open questions. For instance,  we
intend to discuss elsewhere the extension of this explicit study to
the other  black holes. It would  be of interest to  study the
effect of non trivial backgrounds  on black hole spectrums.
Moreover, four and higher  dimensional black holes have been
extensively discussed in connection with string theory and related
topics including attractor mechanism  associated with the Calabi-Yau
compactifications\cite{Vafa,adil}.  It  will be  then an interesting
task to investigate such  black solutions embedded  in non trivial
supergravity theories.   We believe that these complimentary methods
deserve to be studied further.


\begin{thebibliography}{10}

\bibitem{a0} G. W. Gibbons,  S. W. Hawking, {\em Action integrals and partition functions
 in quantum gravity},  Phy. Rev. {\bf D15}  (1977)2752.

\bibitem{a13} S. Weinberg, {\em Gravitation and Cosmology: Principles
and Application of the General Theory of Relativity}, by John Wiley
\& Sons, New York, USA, 1972.

\bibitem{w1} D. N. Page, {\em Hawking radiation and black hole thermodynamics}, New Journal of Physics {\bf 7} (2005) 203.

\bibitem{w2} C. Rong-Gen, {\em Gauss-Bonnet black holes in AdS spaces}, Phy. Rev. D {\bf 65} (2002) 084014.

\bibitem{w3} M. M. Caldarelli, G. Cognola, D. Klemm. {\em Thermodynamics of Kerr-Newman-AdS black holes and conformal field theories},
Clas. Quan. Grav. {\bf 17} (2000) 399.
 \bibitem{w30} D. Kastor, S. Ray and J. Traschen,  {\em Enthalpy and the Mechanics of AdS
Black Holes}, Class.Quant.Grav.{\bf 26}(2009)195011,  {\tt arXiv:0904.2765 [hep-th]}.

\bibitem{w4} A. Belhaj, M. Chabab, H. El Moumni and M. B. Sedra, {\em  On thermodynamics of AdS
black holes in arbitrary dimensions}, Chin. Phys. Lett. {\bf 29}
(2012) 100401.

\bibitem{w5} A. Belhaj, M. Chabab, H. El Moumni, L. Medari and M. B. Sedra,
{\em The thermodynamical behaviors of Kerr–Newman AdS black
holes}, Chin. Phys. Lett. {\bf30} (2013) 090402.

\bibitem{w6} A. Belhaj, M. Chabab, H. El Moumni, K. Masmar and M. B. Sedra, {\em  Critical behaviors
of 3D black holes with a scalar hair}, Int. J. Geom. Meth. Mod.
Phys. {\bf 12}(02) (2014) 1550017.

\bibitem{w7} A. Belhaj, M. Chabab, H. El moumni, K. Masmar and M. B. Sedra, {\em  Maxwell's equalarealaw
for Gauss-Bonnet-Anti-de Sitter black holes}, Eur. Phys. J. C {\bf
75}(2) (2015) 71.

\bibitem{w8} A. Belhaj, M. Chabab, H. El Moumni, K. Masmar, M. B. Sedra and A. Segui, {\em  On
heat properties of AdS black holes in higher dimensions}, J. High
Energy Phys. {\bf 05} (2015) 149.

\bibitem{w9} A. Belhaj, M. Chabab, H. El Moumni, K. Masmar and M. B. Sedra, {\em  On thermodynamics
of AdS black holes in M-theory}, Eur. Phys. J. C {\bf 76}(2) (2016)
73.

\bibitem{w10} M. Chabab, H. El Moumni and K. Masmar,  {\em On thermodynamics of charged AdS
black holes in extended phases space via M2-branes background}, Eur.
Phys. J. C {\bf 76}(6) (2016) 304.

\bibitem{w11} M. Chabab, H. El Moumni, S. Iraoui and K. Masmar,  {\em Behavior of quasinormal modes
and high dimension RN AdS black hole phase transition}, Eur. Phys.
J. C {\bf 76}(12) (2016) 676.

\bibitem{w12} H. El Moumni,  {\em Phase transition of AdS black holes with non linear source in the
holographic framework}, Int. J. Theor. Phys. {\bf 56}(2) (2017)
554-565.

\bibitem{w13} M. Chabab, H. El Moumni, S. Iraoui and K. Masmar, {\em  Phase transition of charged-
AdS black holes and quasinormal modes: A time domain analysis},
Astrophys. Space Sci. {\bf 362} (2017) 192.

\bibitem{w130}A. Belhaj, H. El Moumni, {\em  Entanglement Entropy and Phase Portrait of
f(R)-AdS Black Holes in the Grand Canonical Ensemble}, Nuc. Phys. B{\bf938}(2019) 200.


\bibitem{a1}  R. Fardon, A.  E. Nelson, N.  Weiner,  {\em dark energy from mass varying neutrinos},  Journal
 of Cosmology and Astroparticle Physics {\bf 10} (2004) 005.

\bibitem{a2} E. J. Copeland, M. Sami, S.  Tsujikawa, {\em  Dynamics of dark energy},  International
Journal of Modern Physics {\bf D15} (2006)1753.


\bibitem{ade1} D. N. Spergel et al., { \em Wilkinson Microwave Anisotropy Probe (WMAP) Three Year Results: Implications for Cosmology}, {\tt arXiv:astro-ph/0603449}.

\bibitem{ade2} M. Tegmark et al., { \em Cosmological constraints from the SDSS luminous red galaxies}, Phys. Rev. {\bf D74}(2006)123507 .

\bibitem{w18} S. D. H. Hsu, {\em Entropy bounds and dark energy},  Phys. Lett. B {\bf 594} (2004)13.

\bibitem{w19} R. R. Caldwell, M. Kamionkowski, N. N. Weinberg, {\em Phantom energy: dark energy with $w < 1$ causes a cosmic doomsday},
 Phys.  Rev.  Let. {\bf 91} (2003) 071301.

\bibitem{Chabab:2017xdw}
  M.~Chabab, H.~El Moumni, S.~Iraoui, K.~Masmar and S.~Zhizeh,
  {\em More Insight into Microscopic Properties of RN-AdS Black Hole Surrounded by Quintessence via an Alternative Extended Phase Space},
  Int.\ J.\ Geom.\ Meth.\ Mod.\ Phys.\  {\bf 15}, no. 10, (2018)1850171.



\bibitem{Li:2014ixn}
  G.~Q.~Li,
  {\em Effects of dark energy on P–V criticality of charged AdS black holes},
  Phys.\ Lett.\ B {\bf 735}, (2014) 256, { \tt arXiv:1407.0011 [gr-qc]}.
  \bibitem{q1}  V. V. Kiselev, {\em  Quintessence and black holes}, Class. Quant. Grav. {\bf 20}(2003) 1187, {\tt
gr-qc/0210040}.
	\bibitem{anas1} S. Chen,  J. Jing, { \em Quasinormal modes of a black hole surrounded by quintessence}, Class. Quant. Grav. {\bf 22} (2005)4651, {\tt gr-qc/0511085}.
	
	\bibitem{anas2}  S. Chen , B. Wang, R. Su, {\em Hawking radiation in a d-dimensional static spherically symmetric black hole surrounded by quintessence}, Phys. Rev. D  {\bf 77}(2008) 124011,  {\tt  arXiv: 0801.2053}.
	
   \bibitem{q2} G. G.  Sushant, {\em  Ghosh Rotating black hole and quintessence},  Eur. Phys. J. C {\bf 76} (2016), {\tt  arXiv:1512.05476}.
    \bibitem{q3} Z. Xu, J. Wang, {\em Kerr-Newman-AdS Black Hole In Quintessential Dark Energy},  Phys.Rev. D{\bf 95} (2017)  064015, {\tt  arXiv:1609.02045 }.
     \bibitem{q4} E. Spallucci and A. Smailagic, {\em  Maxwell’s equal area law and the Hawking-Page phase transition}, J. Grav. 2013 (2013) 525696, {\tt arXiv:1310.2186}.
      \bibitem{q5} S. Chen, B. Wang and R. Su, {\em Hawking radiation in a d-dimensional static sphericallysymmetric black Hole surrounded by quintessence}, Phys. Rev. D{\bf 77}, (2008) 124011, {\tt arXiv:0801.2053 [gr-qc]}.
       \bibitem{q6} H. Liu and X. H. Meng, {\em Effects of dark energy on the efficiency of charged AdS black holes as
heat engine}, {\tt  arXiv:1704.04363 [hep-th]}.
        \bibitem{q7} J. Schee, Z. Stuchlik,  {\em Silhouette and spectral line profiles in the special modification of the Kerr black hole geometry generated by quintessential fields},  Eur. Phys. J. C{\bf 76} (2016)643, {\tt arXiv:1606.09037 [astro-ph.HE]}.
         \bibitem{q8} M. Azreg-A\"{\i}nou and M. E. Rodrigues, Thermodynamical, geometrical and Poincar\'e methods
for charged black holes in presence of quintessence, JHEP 1309, (2013) 146, {\tt  arXiv:1211.5909
[gr-qc]}.
          \bibitem{q9}  Y. H. Wei and Z. H. Chu, {\em Thermodynamic properties of a Reissner-Nordstroem quintessence
black hole}, Chin. Phys. Lett. {\bf 28} (2011)100403.
           \bibitem{q10} M. S. Ma, R. Zhao and Y. Q. Ma,  {\em Thermodynamic stability of black holes surrounded by
quintessence}, Gen. Rel. Grav. {\bf 49} (2017) 79,  {\tt  arXiv:1606.06070 [gr-qc]}.




\bibitem{w14}  H. Dong, Q. Y. Cai , X. F. Liu, et al. Commun. Theor. Phys {\bf 61},  (2014) 289.

\bibitem{w15} H. Tasaki, {\em From Quantum Dynamics to the Canonical Distribution: General Picture and a Rigorous Example}, Phys. Rev. Lett.  {\bf 80} (1998) 1373.

\bibitem{w16} S. Goldstein, J. L. Lebowitz, R. Tumulka, and N. Zanghï, {\em Canonical Typicality}, Phys. Rev. Lett {\bf 96} (2006) 050403.

\bibitem{w17} S. Popescu, A. J. Short, and A. Winter, {\em The foundations of statistical mechanics from entanglement: Individual states vs. averages}, Nat. Phys. {\bf 2} (2006) 758.

\bibitem{a4} Y. Ma, J. Chen, and C. Sun, {\em Dark information of black hole radiation raised by
 dark energy},  Nuc. Phys.  B {\bf 931} (2018) 418.

\bibitem{A}  Q. Y. Cai, C. P. Sun, L. You, Nucl. Phys. B {\bf 905} (2016) 327-336 .

\bibitem{key-sw1} S.W. Hawking, {\em Black hole explosions}, Nature {\bf 30} (1974) 248.

\bibitem{key-sw2} S.W. Hawking, {\em Particle creation by black holes}, Commun. Math. Phys {\bf 43} (1975) 199.

\bibitem{a5} Y. H. Ma1,  Q-Y. Cai, H. Dong,  and C-P. Sun, {\em Non-thermal radiation of black holes off canonical
 typicality}, EPL  {\bf 122} (2018) 30001.

\bibitem{QT1} S. Hemming, E. Keski-Vakkuri, {\em Hawking radiation from AdS black holes}, Phys. Rev. D {\bf 64}
 (2001) 044006.

\bibitem{QT2} R. Schützhold and W. G. Unruh, {\em Gravity wave analogues of black holes}, Phys. Rev. D {\bf 66}(2002) 124009.

\bibitem{QT3} M. Alves, Int. J. Mod. Phys. D {\bf 10} (2001) 575.

\bibitem{QT4} E. C. Vagenas, Mod. Phys. Lett. A {\bf 17}(2002) 609 .

\bibitem{QT5} E. C. Vagenas, Phys. Lett. B {\bf 533}(2002) 302.

\bibitem{QT6} E. C. Vagenas, Mod. Phys. Lett. A {\bf 20}(2005) 2449 .

\bibitem{QT7} M. Arzano, A.J.M. Medved, E. C. Vagenas, {\em Hawking Radiation as Tunneling through the Quantum Horizon}, J. High Energy Phys. {\bf 0509} (2005) 037.

\bibitem{X} M. K. Parikh, F. Wilczek,  {\em Hawking radiation as tunneling}, Phys. Rev. Lett. {\bf 85} (2000) 5042.

\bibitem{darkenergymetric} H. J. He and Z. Zhang, JCAP 1708 {\bf 17} (2017) 036.

\bibitem{Y} J. Pu, Y.  Han, {\em Hawking radiation of charged particles via tunneling from the
 Reissner-Nordstrom black hole},  Int. Jour.  of  Theo. Phys. {\bf 56} (2017) 2485.


\bibitem{A1} C. D. Robinson,  {\em Uniqueness of the Kerr black hole},  Phys. Rev.  Let.  {\bf 34} (1975) 905.

\bibitem{a8} B. Toshmatov, Z.  Stuchlik, B.  Ahmedov, {\em Rotating black hole solutions with quintessential
energy},   European Physical Journal Plus {\bf 132} (2017) 98.

\bibitem{a6}  M. Reinhard, {\em Black holes: A physical route to the Kerr metric}, Annalen der Physik {\bf 11}(2002) 509-521.

\bibitem{a7} Z. Jingyi,  Z. Zhao,  {\em Hawking radiation via tunneling from Kerr black holes}, Mod.  Phys. Lett.  A {\bf 20} (2005)1673.

\bibitem{a10} E T. Newman, R. Couch, K. Chinnapared, A. Exton, A. Prakash, R. Torrence, {\em Metric of a rotating, charged mass}, JMP {\bf 6} (1965) 918-919.

\bibitem{a11} C. Doran, {\em New form of the Kerr solution},  Phys.  Rev. D {\bf 61} (2000) 067503.


\bibitem{a9} J. Qing-Quan,  S. Q. Wu,  X. Cai, {\em Hawking radiation as tunneling from the Kerr and Kerr-Newmann black holes}, Phys. Rev.  {\bf D73} (2006) 064003.


\bibitem{a12}  R. Kerner, and Robert B. Mann, {\em Charged fermions tunneling from Kerr–Newmann black holes},
Phys. Lett.  B{\bf665} (2008)277.

\bibitem{Vafa} H. Ooguri, A. Strominger, C. Vafa, {\em Black Hole Attractors and the Topological String}, Phys. Rev.  D{\bf
70} (2004) 106007.

\bibitem{adil} A. Belhaj, {\em On Black Objects in Type IIA Superstring Theory on Calabi-Yau Manifolds}, African Journal of Mathematical Physics, {\bf 6} (2008) 49, {\tt  arXiv:0809.1114}.

\end{thebibliography}
\end{document}